\documentclass[aps,amsfonts,amsmath,prd,preprint,nofootinbib]{revtex4}
\newcommand{\beq}{\begin{equation}}
\newcommand{\eeq}{\end{equation}}

\def\be{\begin{equation}}
\def\ee{\end{equation}}
\def\beqa{\begin{eqnarray}}
\def\eeqa{\end{eqnarray}}

\usepackage{epsfig}

\begin{document}

\title{Holographic multiverse and the measure problem}

\author{Alexander Vilenkin}

\affiliation{Institute of Cosmology, Department of Physics and Astronomy,\\ 
Tufts University, Medford, MA 02155, USA}

\begin{abstract}

We discuss the duality, conjectured in earlier work, between the wave function of the multiverse and a 3D Euclidean theory on the future boundary of spacetime.  In particular, we discuss the choice of the boundary metric and the relation between the UV cutoff scale $\xi$ on the boundary and the hypersurface $\Sigma$ on which the wave function is defined in the bulk.  We propose that in the limit $\xi \to 0$ this hypersurface should be used as the cutoff surface in the multiverse measure.  Furthermore, we argue that in the inflating regions of spacetime with a slowly varying Hubble rate $H$ the hypersurfaces $\Sigma$ are surfaces of constant comoving apparent horizon (CAH). Finally, we introduce a measure prescription (called CAH+) which appears to have no pathological features and coincides with the constant CAH cutoff in regions of slowly varying $H$.  
\end{abstract}

\maketitle

\section{Introduction}

One of the most intriguing unresolved problems of inflationary cosmology is that of calculating probabilities in models of eternal inflation.  The crux of the problem is that the numbers of all kinds of events occurring in the course of eternal inflation are infinite.  Whatever cutoff method is used, most of the events occur close to the cutoff, and the resulting probability measure depends sensitively on the cutoff prescription.

A number of different measures have been proposed and their properties have been investigated.  (For recent discussion and references, see, e.g., \cite{DeSimone:2008if,LVW,BFLR10}.)  This work has shown that some of the proposals lead to paradoxes or to a glaring conflict with the data and should therefore be discarded.  It seems unlikely, however, that this kind of phenomenological analysis will result in a unique prescription for the measure. 

A more satisfactory approach would be to motivate the choice of measure from some fundamental theory.  In this spirit, it was proposed in \cite{holographic1,holographic2} that the wave function of the inflationary multiverse has a dual description in the form of a lower-dimensional Euclidean theory defined at the future boundary ${\cal E}$ of the inflating part of spacetime.  The measure of the multiverse can then be related to the short-distance (UV) cutoff $\xi$ in that theory.  
%It was argued in \cite{holographic1,holographic2} that in the limit $\xi\to 0$ the boundary theory becomes conformally invariant, approaching a UV fixed point.    
This proposal was inspired by the holographic ideas (for a review see \cite{Susskind00,Bousso}) and in particular by the AdS/CFT \cite{Maldacena98,Gubser:98,Witten98} and dS/CFT \cite{Strominger1,Strominger2} correspondence.

Even without knowing the specific form of the boundary theory, one can try to deduce some properties of the resulting measure.   Some attempts in this direction have been made in Refs.~\cite{holographic1,holographic2,Bousso09,BFLR}, but the analysis so far has been inconclusive.  
In the present paper we shall further explore this approach, building mostly on the ideas in \cite{holographic2,BFLR}.  

We begin in the next Section by reviewing the general idea of the boundary-bulk correspondence in the multiverse.  In Section III this correspondence is analyzed in greater detail in the case of a de Sitter universe; in particular, we discuss the relation between the UV cutoff scale $\xi$ on the boundary and the hypersurface $\Sigma$ on which the wave function is defined in the bulk. 
In Section IV the analysis is extended to an inflating universe with a slowly varying expansion rate $H({\bf x},t)$.  Our measure proposal is introduced in Section V.  We discuss the conformal gauge condition which is necessary to fix the boundary metric and suggest that the hypersurfaces $\Sigma$ related to the boundary cutoff $\xi$ should be used as cutoff surfaces in the bulk.  
For regions of slowly varying $H$ we find that these hypersurfaces are the surfaces of constant comoving apparent horizon (CAH).  In Section VI we formulate a simple measure prescription (called CAH+) which agrees with the constant CAH cutoff in regions of slowly varying $H$ and seems to have no pathological features.  The hope is that this prescription may be useful as a simple model of the measure until further progress is made.  Our conclusions are briefly summarized and discussed in Section VII.

\section{The boundary-bulk correspondence}

The spacetime of the inflationary multiverse includes bubbles of all
possible types which nucleate and expand in the inflating background.
The future boundary of this spacetime is comprised of the big crunch 
singularities of the negative-energy
anti-de Sitter (AdS) bubbles, "hats" corresponding to future null and
timelike infinities of Minkowski bubbles, and the eternal set ${\cal
  E}$, which includes spacelike future boundaries of the inflating de
Sitter (dS) bubbles -- or rather what remains of these boundaries
after removal of the regions eaten up by the terminal (that is, AdS
and Minkowski) bubbles.  The boundary theory lives on the eternal set
${\cal E}$. 

If the inflating region of spacetime admits a conformal
treatment of infinity,  the bulk metric in that region can
be represented as  
\beq
g_{\mu\nu}(\eta,{\bf x}) = \Omega^2(\eta,{\bf x}){\tilde g}_{\mu\nu}(\eta,{\bf x}),
\label{gOmegag}
\eeq
where ${\cal E}: \eta=0$ corresponds to the future infinity,
$\Omega(\eta\to 0,{\bf x})\to\infty$, while ${\tilde g}_{\mu\nu}$
remains finite.  In this case,
${\cal E}$ can be identified with the null and timelike future infinities
of the inflating region,  
\beq
{\cal E}={\mathcal I}_+= i_+.
\eeq
The asymptotic form of the metric, 
\beq
{\bar g}_{ij}({\bf x})={\tilde g}_{ij}(0,{\bf x}) ,
\eeq
can be used as the metric on ${\cal E}$.  If the universe is closed, 
${\bar g}_{ij}$ is a metric on a Euclidean compact manifold.  
%We shall assume this to be the case in the present paper.
The assumption that infinities in the asymptotic metric can be
factored out as in Eq.(\ref{gOmegag}) is probably too strong and may
be unnecessary, but we shall adopt it here to simplify the discussion.

The duality relation suggested in \cite{Maldacena03,holographic1,holographic2} is
\beq
\Psi [{\bar\phi}({\bf x})] \equiv \int D\phi e^{iS[\phi]} = e^{iW[{\bar\phi}]}.
\label{duality}
\eeq
Here, $S$ is the bulk action and the integral is over bulk fields
$\phi$ (including the metric) approaching the prescribed values ${\bar\phi}({\bf x})$ at the
boundary.  These values play the role of external sources for the
fields of the boundary theory.  The amplitude $\Psi[{\bar\phi}({\bf
    x})]$ has the meaning of the asymptotic wave function of the
multiverse, and $W[{\bar\phi}]$ is the effective action for the boundary
theory with appropriate couplings to the sources ${\bar\phi}$.

Because of the infinite conformal factor, any finite wavelength on the
boundary ${\cal E}$ corresponds to an infinite asymptotic wavelength
in the bulk.  For example, the wavelengths of gravitons produced as
quantum fluctuations in de Sitter bubbles grow unboundedly with time.  Once
the wavelength becomes much greater than the horizon, the metric
perturbation `freezes', and its behavior becomes increasingly
classical, approaching precise classicality in the limit \cite{GuthPi}.
Light scalar fields with masses $m<3H/2$, where $H$ is the de Sitter
expansion rate, exhibit similar behavior, except the amplitudes of the
waves scale like $\Omega^{-\beta}$, with the conformal weight $\beta$
depending on the mass.  The asymptotic light fields ${\bar\phi}({\bf
  x})$ are thus classical fields whose particular configuration
represents a frozen history of quantum fluctuations in the inflating
region.  Massive fields with $m>3H/2$ have complex conformal weights
\cite{Strominger1} and do not approach classicality.  We assume that
the scaling with $\Omega$ has been factored out in the definition of
the fields ${\bar\phi}({\bf x})$ in (\ref{duality}).   
  
Once we have a metric on ${\cal E}$, we can impose an ultraviolet (UV)
cutoff in the boundary theory at some distance scale $\xi$.   The cutoff smoothes out all perturbations on scales smaller than $\xi$ and removes all bubbles of size less than $\xi$ at the boundary.  By analogy with AdS/CFT correspondence, it was suggested in \cite{holographic1,holographic2} 
that the boundary theory with a UV cutoff is dual to the bulk theory
with an infrared (IR) cutoff on a "late-time" spacelike hypersurface.
This conjectured duality can be symbolically expressed as 
\beq
\Psi_\Sigma [{\bar\phi}({\bf x})] = e^{iW_\xi [{\bar\phi}]}.
\label{duality2}
\eeq
Here, $W_\xi [{\bar\phi}]$ is the boundary effective action with a UV
cutoff $\xi$ and  
$\Psi_\Sigma [{\bar\phi}({\bf x})]$ is the wave function evaluated on
a finite spacelike hypersurface $\Sigma$.  The relation between $\xi$
and $\Sigma$ will be discussed below, but generally we expect that $\Sigma\to
{\cal E}$ as $\xi\to 0$.  This suggests that the renormalization group
(RG) flow in the boundary theory corresponds to time evolution in the
bulk \cite{Strominger2} (see also \cite{VS,LMN}).  

The conformal factor $\Omega(\eta,{\bf x})$ in (\ref{gOmegag}) is
defined up to finite rescalings,  
\beq
\Omega(\eta,{\bf x})\to f(\eta,{\bf x})\Omega(\eta,{\bf x}),
\eeq
with $f(\eta,{\bf x})<\infty$, which correspond to finite Weyl transformations of the metric on ${\cal E}$,
\beq
{\bar g}_{ij}({\bf x})\to {\bar f}^2({\bf x}){\bar g}_{ij}({\bf x})
\label{Weyl}
\eeq
with ${\bar f}({\bf x})=f(0,{\bf x})$.  As we shall discuss in Section III.D, different boundary metrics in this conformal class correspond to different choices of the hypersurfaces $\Sigma$ in the bulk spacetime.\footnote{The boundary action $W$ is not generally invariant under Weyl rescalings, due to the conformal anomaly.  This is consistent with the fact that the wave function $\Psi$ is not invariant under variation of the hypersurface $\Sigma$.  There are no conformal anomalies in $3D$, but anomalies will still be present at the $2D$ boundaries between different vacua.  I am grateful to Jaume Garriga for pointing this out to me.}

\section{The UV/IR correspondence in de Sitter}

\subsection{dS/CFT}

To determine the relation between the boundary cutoff $\xi$ and the surface
$\Sigma$, we shall first consider the case of a single de Sitter vacuum.
The unperturbed metric is then 
\beq
ds^2 = a^2(\eta)(-d\eta^2 + d{\bf x}^2)
\label{dS}
\eeq
with
\beq
a(\eta)=(H|\eta|)^{-1}.
\eeq
The future infinity is at $\eta = 0$, with the metric
\beq
d{\bar s}^2=d{\bf x}^2. 
\label{sbarflat}
\eeq
We will be interested in the short-distance behavior at late times, that is, in the regime where $|\Delta {\bf x}|\to 0$ and $\eta\to 0$.  In this regime, the global structure of spacetime is unimportant, so the flat slicing (\ref{dS}) of de Sitter can be used without loss of generality.

The action for a de Sitter universe with a future boundary on a surface of 
$\eta=const$ is (in reduced Planck units $8\pi G=1$)
\beq 
S_0=-3\int d^3 x\int^\eta d\eta' \left[\left(\frac{da}{d\eta'}\right)^2 +H^2 a^4\right] =-6{\mathcal V}H^{-2} \int^\eta \frac{d\eta'}{{\eta'}^4},
\label{S0int}
\eeq
where
\beq
{\mathcal V} = \int d^3 x
\eeq
is the integrated comoving volume.
In the late time limit, $\eta\to 0$, 
\beq 
S_0 = -2{\mathcal V} H a^{3}.
\label{W0}
\eeq
We can think of $\Psi_0 = \exp(iS_0)$ as the wave function of a compact universe of total comoving volume ${\mathcal V}$.  Alternatively, we can interpret it as the wave function for a fixed comoving volume in a larger universe.  

The contribution of linearized gravitons to the wave function $\Psi$ has been studied in Refs.~\cite{Maldacena03} and \cite{holographic2}.  Up to quadratic order in perturbations, it can be expressed as $\exp(iS_2)$, where $S_2$ is given by
\beq
S_2 [{\bar h}({\bf x})] = \frac{1}{2H^2} \sum_{\bf k} \left(-Hak^2 +ik^3 +{\cal O}(\eta) \right)|h_{\bf k}|^2.
\label{S2}
\eeq
Here, $h({\bf x})$ is the amplitude of transverse traceless metric perturbations, and
\beq
h_{\bf k}=\frac{1}{\sqrt{\cal V}}\sum_{\bf k} e^{-i{\bf kx}}h({\bf x}).
\eeq

The zeroth-order term (\ref{W0}) and the term proportional to $\eta^{-1}$ in (\ref{S2}) diverge as $\eta\to 0$.  These terms can be identified with the cubic and linearly divergent terms in the effective action of a $3D$ Euclidean field theory, 
\beq
W_{div} = \int_{\mathcal V} d^3 x \sqrt{g} (b_3 \xi^{-3} + b_1 \xi^{-1} R), 
\label{WCFT1}
\eeq
where as before $\xi$ is the UV cutoff length.  
The form of the finite, non-analytic term in (\ref{S2}), 
\beq
W_{fin} = b_0 \sum_{\bf k} k^3 |h_{\bf k}|^2  = b_0 \int_{\mathcal V} d^3 x h({\bf x}) (\partial^2)^{3/2} h({\bf x}),
\label{WCFT2}
\eeq
is consistent with Weyl invariance of the boundary theory \cite{Duff,Marolf}.  (Note that there are no conformal anomalies in odd number of dimensions.)  In the bulk, this term describes a scale-invariant spectrum of gravitons.

The coefficients $b_0$, $b_1$ and $b_3$ are expected to be proportional to the comoving volume ${\mathcal V}$ and, assuming no major cancellations, to the number of fields ${\mathcal N}_b$ (central charge) of the boundary CFT.  Then, comparing Eqs.~(\ref{WCFT1}), (\ref{WCFT2}) with (\ref{W0}),(\ref{S2}), it is easy to see that in order to have $\Psi = e^{iS}=e^{iW}$, we should make the identifications
\beq
\xi \sim (Ha)^{-1}=|\eta| ,
\label{xia}
\eeq
\beq
{\mathcal N}_b \sim H^{-2}.
\label{NbH}
\eeq
The meaning of the relation (\ref{xia}) is very simple.  Rewriting it as $a(\eta)\xi \sim 1/H$, it says that the comoving cutoff scale $\xi$ corresponds to the horizon scale $H^{-1}$ on the surface $\Sigma$.

\subsection{Counting degrees of freedom}

In order for our conjectured duality to hold, the bulk and boundary theories should have the same number of field degrees of freedom \cite{SW98}.  Let us now check if this is indeed the case.

The regulated boundary theory is essentially a discrete theory, like a lattice field theory, with a characteristic lattice spacing equal to $\xi$.  The number of field degrees of freedom per one lattice site is equal to the number of independent fields in the theory,  Eq.~(\ref{NbH}).
This number is ${\mathcal N}_b \gg 1$ if the de Sitter expansion rate $H$ is much smaller than the Planck scale, $H\ll 1$.
As we have just discussed, the cutoff scale $\xi$ on the boundary corresponds to the horizon scale $H^{-1}$ on the surface $\Sigma$ in the bulk.  If we subdivide $\Sigma$ into horizon-size cells, the duality suggests that each cell should carry $\sim {\mathcal N}_{b}$ degrees of freedom.

The entropy in a horizon-size region is $S_h \sim S_{GH}$, where $S_{GH}\sim H^{-2}$ is the Gibbons-Hawking entropy.  Entropy cannot exceed the number of degrees of freedom in the region ${\mathcal N}_h$; hence\footnote{The number of degrees of freedom is defined as the logarithm of the dimension of the corresponding Hilbert space.}  
\beq
{\mathcal N}_h\gtrsim H^{-2}. 
\label{N<H}
\eeq
On the other hand, according to the holographic principle \cite{Susskind00,Bousso}, the number of bulk degrees of freedom in a region enclosed by a surface of area $A$ is bounded by
\beq
{\mathcal N} \leq \frac{1}{4} A.
\label{Nbound}
\eeq
For a horizon-size region this gives ${\mathcal N}_{h}\lesssim H^{-2}$. Combining this with (\ref{N<H}), we conclude that
\beq
{\mathcal N}_h\sim H^{-2}, 
\label{N=H}
\eeq
and thus \cite{holographic2}
\beq
{\mathcal N}_h\sim{\mathcal N}_b ,
\label{NhNb}
\eeq
as expected from the duality.

The field degrees of freedom in different cutoff-size regions in the boundary theory are completely independent, which implies that the same should be true for the degrees of freedom in different horizon regions on $\Sigma$.  This is somewhat unorthodox, but this picture appears to be consistent with the holographic ideas.

An apparent inconsistency would occur if we try to apply the bound (\ref{Nbound}) to a region of size $L\gg H^{-1}$.  This would give ${\mathcal N}\lesssim L^2$, while our picture suggests ${\mathcal N}\sim (LH)^3 H^{-2} \sim HL^3 \gg L^2$.  In fact, it has long been recognized that the holographic bound (\ref{Nbound}) and the related entropy bound
\beq 
S\leq\frac{1}{4} A
\label{Sbound}
\eeq
(which follows from (\ref{Nbound})) do not apply in the cosmological context on a sufficiently large scale $L$ \cite{FischlerSusskind}.  Bousso \cite{Bousso1,Bousso2} presented strong evidence that the bound (\ref{Sbound}) is generally applicable if $S$ is interpreted as the entropy on a light sheet emanating from a $2D$ surface ${\mathcal A}$ of area $A$.
A light sheet is defined as a congruence of null geodesics which is orthogonal to the surface and is converging (not diverging in the marginal case) at all points of the surface.  As long as the null energy condition is satisfied, the geodesics will continue to converge after they leave the surface ${\mathcal A}$, until they cross at a caustic in a finite affine parameter time.   If the null geodesics are terminated when they cross another geodesic of the congruence, then the light sheet is a compact surface and ${\cal A}$ is its only boundary.  A future-directed light sheet is to the future of any spacelike surface bounded by ${\cal A}$.  Thus the entropy on such a surface is bounded by the entropy on the light sheet, which is in turn bounded by $A/4$ \cite{Bousso1,Bousso2}.
It follows that the bound (\ref{Sbound}) applies to any $3D$ region on $\Sigma$ which is bounded by a surface ${\mathcal A}$ admitting a future-directed light sheet.  This only holds for regions of sub-horizon size, $L<H^{-1}$.  

Bousso's analysis does not tell us anything about the number of degrees of freedom in super-horizon regions.  Our duality picture fills this void, suggesting that the degrees of freedom in different horizon regions are independent.

\subsection{Time evolution and unitarity}

The UV/IR relation (\ref{xia}) indicates that the RG flow in the boundary theory corresponds to time evolution in the bulk, as proposed in \cite{Strominger2}.  We shall now discuss some unusual features of this correspondence.

Our picture suggests that sub-horizon and super-horizon scales in the bulk are represented in radically different ways in the boundary theory.  On super-horizon scales, massless and light fields are "frozen in" with the expansion and are propagated to future infinity by conformal stretching.  
The field pattern is then imprinted on ${\cal E}$ and is retrieved in the regulated theory with a resolution depending on the boundary cutoff $\xi$.  The correspondence between the bulk and boundary fields in this regime is thus very direct.  

On the other hand, the physics of each horizon region on $\Sigma$ is encoded in the ${\cal N}_{b}\sim H^{-2}$ field degrees of freedom in the corresponding cutoff-size region on ${\cal E}$.  This should give a rather scrambled representation: 
deciphering the quantum state of sub-horizon regions from the boundary theory may be nearly as difficult as deciphering the physics of black hole interiors from Hawking radiation.  We note that the situation here is similar to that in AdS/CFT correspondence, where objects larger than the AdS radius $R_{AdS}$ have a simple geometric representation in the boundary theory, while the representation of objects smaller than $R_{AdS}$ is far from straightforward and is not yet fully understood \cite{SW98}.

Another intriguing aspect of our picture is that the number ${\mathcal N}_b \sim H^{-2}$ of field degrees of freedom in a cutoff-size cell on the future boundary is sufficient for a complete description of the corresponding horizon region in the bulk.  
At the same time, the cutoff eliminates all degrees of freedom with wavelengths shorter than $\xi$ on the boundary.  As the cutoff length $\xi$ is increased, more and more boundary degrees of freedom are integrated out.  In the bulk this corresponds to moving the surface $\Sigma$ backwards in time.  The number of horizon regions per comoving volume on $\Sigma$ is then decreased, and the number of field degrees of freedom is decreased accordingly.

Time evolution in this picture cannot be described by a unitary operator, as usual in Quantum Mechanics.  The situation is especially clear cut if we assume that the universe originates as a closed, horizon-size 3-sphere, as suggested in \cite{AV82}.  Then, at the time of nucleation, the number of degrees of freedom is ${\mathcal N}\sim H^{-2}$.  This number grows unboundedly in the course of subsequent expansion.  As we go from later to earlier times, short wavelength modes are integrated out, so information is being lost, while we know that information loss does not occur in the course of unitary evolution of the wave function in Quantum Mechanics.

This is not to say that time evolution is non-unitary in the sense that probabilities do not add up to one.  Of course they do.  The evolution backwards in time is equivalent to coarse-graining, where the probabilities of finer-resolution configurations are lumped together to yield the probabilities of coarse-grained configurations.\footnote{We note in passing that absence of unitarity is not necessarily problematic in the context of quantum cosmology.  In the Wheeler-DeWitt approach, the wave function of the universe is time-independent, and it has been argued in \cite{AV89} that time and unitarity emerge as approximate concepts in the semiclassical limit.}

\subsection{Spacetime and Weyl rescalings}

The cosmological wave function $\Psi_\Sigma ({\bar\phi})$ assigns amplitudes to different intrinsic geometries and field configurations on a $3D$ hypersurface $\Sigma$, but does not specify how this hypersurface is embedded in a $4D$ spacetime.  The spacetime continuum is an emergent concept that appears only in the semiclassical limit.\footnote{Once the wave function $\Psi=\exp(iW)$ is found, the semiclassical spacetime can be constructed using the standard technique \cite{Gerlach}.  In the WKB approximation, the functional $W[{\bar\phi}]$ is identified with the classical Hamilton-Jacobi action.  After fixing the gauge by specifying the lapse and shift functions, the spacetime metric can be found by solving the Hamilton-Jacobi equations with this action.} Time evolution can then be described in terms of a foliation of semiclassical spacetime by a one-parameter family of spacelike surfaces.  In our approach, the corresponding time parameter can be identified with the RG flow parameter of the  boundary theory.  

With a flat boundary metric, which we have assumed so far, the foliation surfaces are surfaces of $\eta=const$, Eq.~(\ref{xia}).  We could instead choose a Weyl rescaled metric,
\beq
d{\bar s}^2 = f^2({\bf x}) d{\bf x}^2.
\label{rescaled}
\eeq
If $f({\bf x})$ is slowly varying on the scale $\xi$, $\xi |\nabla f|\ll f$, the boundary effective action should still be given by (\ref{WCFT1}),(\ref{WCFT2}),
\beq
W_{div} = b_3 \xi^{-3} \int_{\mathcal V} d^3 x f^3({\bf x}) + ... 
\label{WCFT3}
\eeq
This can be matched to the bulk action $S_0$ in (\ref{S0int}) on a curved hypersurface $\eta = \eta({\bf x})$ with a slowly varying function $\eta({\bf x})$, 
\beq 
S_0=-6\int_{\mathcal V} d^3 x\int^{\eta({\bf x})} \frac{d\eta'}{{\eta'}^4}
= -2H^{-2}\int_{\mathcal V} d^3 x |\eta({\bf x})|^{-3}.
\label{S0int3}
\eeq
With $b_3\sim H^{-2}$, a comparison of (\ref{WCFT3}) and (\ref{S0int3}) gives 
\beq
\eta({\bf x}) = -\xi/f({\bf x}).
\eeq
This relation indicates that different choices of conformal gauge in the boundary theory correspond to different foliations in the bulk, as it was suggested in \cite{holographic1,holographic2}.

The cutoff length $\xi$ on the boundary corresponds to a coordinate distance $|\Delta {\bf x}| = \xi/f({\bf x}) = -\eta({\bf x})$.  This shows that, as one might expect, the comoving cutoff scale $a(\eta)|\Delta {\bf x}|$ still corresponds to the horizon scale $H^{-1}$ on $\Sigma$.

\section{Extending to the multiverse}

\subsection{Slowly varying $H$}

To go beyond a single dS vacuum, we shall consider a spacetime which is locally de Sitter 
with an expansion rate $H({\bf x},t)$ slowly varying in space and time,
\beq
|\nabla H|\ll H^2.
\label{slowvar}
\eeq 
This is the situation we expect to have on super-horizon scales in models of eternal inflation with quantum diffusion \cite{AV83,Linde86}.  The corresponding spacetime metric can be approximated as
\beq
ds^2\approx a^2({\bf x},\eta)(-d\eta^2+d{\bf x}^2),
\label{dsxeta}
\eeq
with
\beq
a({\bf x},\eta)=\frac{1}{H({\bf x},\eta)|\eta|}.
\label{aHeta}
\eeq
As before, the future boundary of spacetime ${\cal E}$ is at $\eta=0$, and after factoring out the scale factor $a^2$ the boundary metric can be chosen as
\beq
d{\bar s}^2\approx d{\bf x}^2.
\label{sbarflat2}
\eeq

In the regime (\ref{slowvar}), we expect the wave function $\Psi_\Sigma [h] = \exp(iS[h])$ to be given by the same expressions as in Section III.A, except now $H$ and $a$ will be replaced by their local values under the integral over $d^3 x$.  For example, Eq.~(\ref{W0}) will become 
\beq 
S_0 = -2\int d^3 x H({\bf x},\eta) a^{3}({\bf x},\eta),
\label{S_0}
\eeq
with a slowly-varying function $\eta=\eta({\bf x})$ specifying the surface $\Sigma$.

Once again, the duality (\ref{duality}) implies the identification
\beq
\xi \sim [H({\bf x},\eta)a({\bf x},\eta)]^{-1},
\label{xiHx}
\eeq
where $\eta=\eta({\bf x})$.  With $a({\bf x},\eta)$ from (\ref{aHeta}), this becomes
\beq
\xi \sim |\eta|,
\label{xisimeta}
\eeq
which tells us that the hypersurfaces $\Sigma$ are surfaces of $\eta=const$, as before.  

The meaning of Eq.~(\ref{xiHx}) is that the comoving cutoff scale $\xi$ corresponds to the local  apparent horizon scale $H^{-1}({\bf x},\eta)$ on $\Sigma$.  In other words, $\Sigma$ is a surface of constant comoving apparent horizon.  

Different spacetime foliations can be obtained with a different choice of conformal gauge on the boundary.  We note however that some foliations that look simple and natural from the bulk point of view correspond to rather contrived choices of the boundary metric.\footnote{Note that for any choice of $f({\bf x})$ the comoving cutoff scale $|\Delta {\bf x}| = \xi/f({\bf x})$ corresponds to the horizon scale on $\Sigma$, even though $\Sigma$ may no longer be a surface of constant comoving apparent horizon.}  For example, in order to yield a foliation with constant scale factor surfaces, $a=\xi^{-1}$, the function $f({\bf x})$ in (\ref{rescaled}) should satisfy the relation
\beq
f({\bf x})=H\left( {\bf x},\frac{\xi}{f({\bf x})}\right).
\eeq

\subsection{The number of field degrees of freedom}

The analogue of Eq.~(\ref{NbH}) for the number of fields ${\mathcal N}_b$ in the case of a slowly varying $H$ is
\beq
{\mathcal N}_b({\bf x})\sim H^{-2}({\bf x}).
\label{Nbx}
\eeq
The analysis in Sec.~III.B leading to Eq.~(\ref{NhNb}) should still be valid, so we must have
\beq
{\mathcal N}_h \sim {\mathcal N}_b,
\label{NhNb2}
\eeq
where ${\mathcal N}_h$ is the number of field degrees of freedom in a horizon-size region.
As before, we regard this as a consistency check of the proposed duality.

Strictly speaking, our discussion here applies only to spacetimes in which $H$ is slowly varying everywhere.  (We shall call them `slow-$H$ spacetimes' for brevity.)  However, the relations (\ref{Nbx}) and (\ref{NhNb2}) could have a greater generality.  In particular, one could expect them to apply when the surface $\Sigma$ lies in a slow-$H$ region, while the spacetime to the future of $\Sigma$ may not be slow-$H$.  For example, $\Sigma$ could lie in the inflating region which is followed by radiation and matter dominated epochs, like in our part of spacetime.  
In this case, the causal horizon on $\Sigma$ can be much greater than the apparent horizon $H^{-1}$, as evidenced by the fact that we can now observe a multitude of inflationary apparent horizon regions on the CMB sky.  It should then be clear that if the relations (\ref{Nbx}),(\ref{NhNb2}) are to apply, `the horizon' should be understood as the apparent, rather than causal horizon.

A somewhat puzzling aspect of the relation (\ref{Nbx}) is that ${\mathcal N}_b$ is a discrete variable, while the Hubble rate $H$ appears to be continuous.  To address this apparent contradiction, we first note that our approach assumes the validity of the semiclassical approximation, which requires that $H\ll 1$, and thus ${\mathcal N}_b\gg 1$.  In this limit, variation of ${\mathcal N}_b$ in units of $\Delta{\mathcal N}_b\sim 1$ is well approximated as continuous.  We note also the arguments presented in \cite{Mukhanov} indicating that the area of black hole horizon must be quantized in Planck units.  It has also been conjectured \cite{MukhanovAV} that the same applies to the de Sitter horizon.  The horizon area is precisely the quantity on the right hand side of (\ref{Nbx}), and thus horizon area quantization is consistent with Eq.~(\ref{Nbx}).   Regions with different values of ${\mathcal N}_b$ might be separated on ${\cal E}$ by 2-branes, with ${\mathcal N}_b$ changing by one unit across the branes.

\subsection{Nested de Sitter bubbles}

We next consider the model of nested spherical de Sitter bubbles, where the spacetime consists of de Sitter regions of different vacuum energy separated by thin domain walls \cite{holographic2}.  The future boundary  ${\cal E}$ in this case consists of nested spherical regions of different vacua (with bubble collisions being neglected).  The asymptotic $4D$ metric near the boundary $(\eta\to 0)$ can be written in the form (\ref{dsxeta}),(\ref{aHeta}), but now $H({\bf x})$ takes constant values $H_i$ in regions of different vacua (labeled by $i$).  Represented in this way, the metric appears to be discontinuous, but this is because different vacuum regions are not properly matched.  The bulk spacetime is of course continuous; in fact it has been shown in  \cite{holographic2} that it can be foliated by flat hypersurfaces.  As before, the simplest choice of the boundary metric, obtained by factoring out the scale factor $a^2$, is the flat metric (\ref{sbarflat2}).  

A UV cutoff $\xi$ on ${\cal E}$ removes all perturbations of comoving wavelength smaller than $\xi$ and all bubbles of comoving size less than $\xi$.  It also smears the sharp boundaries between different vacuum regions on a distance scale $\sim \xi$.   With a flat boundary metric, the identification
\beq 
\xi \sim (Ha)^{-1}
\eeq
should still apply, except within a comoving distance $\sim\xi$ of the bubble walls.  Apart from these transition regions, the surface $\Sigma$ is then a surface of constant CAH, $\eta=-\xi$, on which the comoving cutoff scale $\xi$ corresponds to the local horizon scale $H_i^{-1}$.

\subsection{Terminal bubbles}

Our discussion has been focused on the inflating region of spacetime, which does not include AdS or Minkowski bubbles.  The proposal of Refs.\cite{holographic1,holographic2} was that the images of such bubbles should be excised from the future boundary and should be represented by Euclidean $2D$ theories living on the boundaries of the resulting "holes".  Further insights into holography of terminal bubbles have been suggested in Refs.~\cite{Maldacena10,Susskind10,Jaume10}, but still it remains unclear how the cutoff surfaces should be continued from the inflating region into the bubble interiors.

\section{The measure proposal}

Turning now to the measure problem, we shall require that hypersurfaces  $\Sigma$ related to the  RG flow on the boundary through the duality (\ref{duality2}) should be used as cutoff surfaces in the bulk.  Note that this prescription is different from the one proposed in Refs.~\cite{holographic1,holographic2}.  

The duality (\ref{duality2}) relates the boundary theory to the wave function {\it on} $\Sigma$, not in the past of $\Sigma$.  Since the measure counts events that occurred prior to the cutoff, one might object to the use of $\Sigma$ as a spacetime cutoff.  Here we take the point of view that we should only count the events for which there is some sort of record on $\Sigma$.  For example, if we want to count Boltzmann brains (BBs) -- the freak observers spontaneously fluctuating out of the vacuum (see, e.g., \cite{DeSimone:2008if} and references therein), we should count only those which left some trace of their existence.  This excludes BBs fluctuating in and out of existence in Minkowski regions of the multiverse, in which case energy conservation forbids any traces of BBs to persist after the fluctuation is over. 

As we discussed in Section III, the RG flow on the boundary corresponds to time evolution in the bulk, and in this sense our measure proposal amounts to simply including the events that occurred prior to a certain time $t$ and taking the limit $t\to\infty$.  In the semiclassical regime, the duality relation provides a foliation of spacetime and thus specifies a time variable.

\subsection{Fixing the conformal gauge}

Different choices of conformal gauge on the boundary give different spacetime foliations and generally yield different probability measures.\footnote{I am grateful to Ben Freivogel and Jaume Garriga for clarifying discussions of this issue and to Ben Freivogel for pointing out an error in the original version of the paper.}   For example, if we perform a Weyl rescaling
\beq
{{\bar g}'}_{ij}({\bf x})= f^2({\bf x}){\bar g}_{ij}({\bf x})
\label{g'g}
\eeq 
with a function $f({\bf x})$ which is correlated with the values of the Higgs fields ${\bar\varphi}_a ({\bf x})$, the probability distribution for different vacua will be changed.  We would like to choose the boundary metric so that the multiverse is sampled in an unbiased way, without giving a preferential treatment to one vacuum over another.  In our holographic picture, we shall interpret this to mean that the boundary metric should be uncorrelated with the local vacua.  More specifically, we shall require that the metric ${\bar g}_{ij}$ is fixed by imposing a gauge condition 
\beq
F({\bar g}_{ij})=0 ,
\label{F=0}
\eeq
where the function $F$ depends only on ${\bar g}_{ij}$ and is independent of matter fields ${\bar\varphi}_a$.  

The function $F$ in (\ref{F=0}) should be generally covariant and have the property that for any metric ${\bar g}_{ij}({\bf x})$ there exists $f({\bf x})$ such that the Weyl transformed metric
(\ref{g'g}) satisfies (\ref{F=0}).   We do not know of any general principle what would fix the form of the function $F$, so we are led to conjecture that all gauge conditions having these properties yield identical probability measures.

In fact, the choice of possible gauge conditions is somewhat restricted.  The function $F$ should be a scalar constructed out of the curvature tensor and its derivatives.  The curvature generally develops delta-function singularities at the bubble walls.\footnote{Delta functions at bubble walls can be removed by a Weyl transformation in the model of nested de Sitter bubbles, but this cannot be done in general, e.g., in the presence of gravitational waves \cite{holographic2}.} 
The delta functions in the scalar curvature ${\bar R}$ can be removed by a Weyl rescaling, but this cannot generally be done for more than one component of the Riemann tensor simultaneously, so products like ${\bar R_{ij}}{\bar R}^{ij}$ are not well defined.  Noticing that according to the Bianchi identity $\nabla_i {\bar R}^{ij}=\frac{1}{2}\nabla^j{\bar R}$, we conclude that $F$ can be a function only of scalar curvature ${\bar R}$ and its derivatives. 

Here we shall adopt the simplest choice of the gauge condition
\beq
{\bar R}=0 ,
\label{R=0}
\eeq
which was first introduced in \cite{holographic2}.   Given a metric ${\bar g}_{ij}({\bf x})$, the function $f({\bf x})$ in (\ref{g'g}) that yields ${\bar R}' =0$ can be found by solving the differential equation
\beq
8\nabla^2 y + {\bar R}y = 0
\label{y}
\eeq
with $y=f^{1/2}$.  We are interested in solutions of (\ref{y}) which are nonsingular and everywhere positive (so that the metric is non-degenerate).  It may not be possible to find such solutions on the entire boundary.  For example, if ${\bar g}_{ij}$ describes a compact space with ${\bar R}>0$, then integration of (\ref{y}) over the boundary volume gives
\beq
\int{\bar R}y\sqrt{\bar g} d^3 x = 0 ,
\eeq
indicating that $y$ cannot be everywhere positive.  It should be possible, however, to cover the boundary with overlapping domains where (\ref{R=0}) can be imposed.  

The gauge condition (\ref{R=0}) still leaves the freedom of Weyl rescaling with $f({\bf x})$ satisfying 
\beq
\nabla^2 f^{1/2} = 0.
\eeq
For example, in models allowing a flat boundary metric, we can also choose
\beq
d{\bar s}^2 = (1+{\bf bx})d{\bf x}^2 
\eeq
with ${\bf b}=const$.  Since any smooth metric can be approximated as flat on sufficiently small scales, this should leave the probability distribution unchanged in the limit $\xi\to 0$.  

More generally, for any smooth function $f({\bf x})$ which does not depend on $\xi$, we have $\xi |\nabla f|/f \to 0$ in the limit $\xi\to 0$, indicating that the characteristic variation scale of $f$ becomes arbitrarily large compared to $\xi$ in this limit.  We shall refer to Weyl rescalings (\ref{g'g}) with such functions $f$ as large-scale Weyl transformations.  Such transformations cannot introduce any correlation of the metric with the local vacua and should therefore leave the probability distribution unchanged.

A simple alternative to the gauge condition (\ref{R=0}), suggested in \cite{BFLR}, is to require 
\beq
{\bar R}=R_0 , 
\label{R=C}
\eeq
where $R_0$ is a positive constant.  It has been shown in \cite{BFLR} that this condition can always be globally satisfied.  Since we are ultimately interested in the limit of small $\xi$, the curvature scale $R_0^{-1/2}$ gets arbitrarily large compared to $\xi$, $R_0^{1/2}\xi\to 0$.  In this limit, the gauge condition (\ref{R=C}) is equivalent to (\ref{R=0}).

For any conformally flat metric, the condition (\ref{R=0}) admits a flat boundary geometry.  This applies in particular to slow-$H$ spacetimes and to the model of nested de Sitter bubbles.   
A flat metric cannot be selected in the presence of gravitational waves.  We note, however, that, for a flat background metric, the condition (\ref{R=0}) still holds if gravitational waves are included as linearized perturbations, as we did in Section III.A.

\section{The CAH+ measure}

In slow-$H$ regions of spacetime, our gauge fixing condition (\ref{R=0}) selects a flat boundary metric, and it follows from the discussion in Sections III,IV that the hypersurfaces $\Sigma$ should be surfaces of constant comoving apparent horizon (CAH).  In order to fully specify the measure, we still need to learn how these surfaces are to be extended to matter and radiation-dominated regions, to terminal bubbles, and to inflating regions where the slow variation condition (\ref{slowvar}) is not satisfied.  Progress in this direction would require a better understanding of the duality, and in particular of how sub-horizon scales are represented in the boundary theory.  Our goal here will be much more modest.  We shall try to formulate a simple measure prescription which has no pathological features and coincides with the constant CAH cutoff in spacetime regions satisfying (\ref{slowvar}).  The hope is that such a prescription will encapsulate what we learned from multiverse holography so far and will be useful as a simple model of the measure until further progress is made.

\subsection{The causality condition}

The first possibility that comes to mind is simply to use constant CAH cutoff surfaces in the entire spacetime.  This can be called the CAH measure.\footnote{A related measure has been recently discussed by Bousso et.~al. in \cite{BFLR10}.  They propose to include all events within the apparent horizon of a single geodesic observer.  With a suitable choice of ensemble of such observers, this measure may be equivalent to CAH.}  As we shall see, however, this measure is problematic, because the constant CAH surfaces are not generally spacelike.  Whatever the rule determining $\Sigma$ eventually turns out to be, we can say with certainty that $\Sigma$ has to be spacelike.   The reason is that the wave function of the universe $\Psi_\Sigma$ in the duality relation (\ref{duality2}) is defined only for spacelike 3-geometries.  The metric components $g_{ij}({\bf x})$ at different points ${\bf x}$ on $\Sigma$ are regarded as independent, commuting quantum variables, which would not be the case for a non-spacelike $\Sigma$.  

We note also that a cutoff surface $\Sigma$ that includes timelike segments is problematic from the point of view of measure phenomenology.    The cutoff surface should generally have the following causality property: if some event $A$ is not removed by the cutoff, then all events that could have influenced $A$ should not be removed as well.  This is a desirable property, since otherwise the measure will assign non-negligible probability to some events in the absence of their causes.  It follows from this criterion that $\Sigma$ should be spacelike or null.

To amend the CAH measure, we shall introduce an additional simple rule which enforces causality.  Given a constant CAH surface $\Sigma$, we shall exclude not only the spacetime region above $\Sigma$, but also the interiors of the future lightcones of all points $P\in\Sigma$.\footnote{A similar amendment (but with a different motivation) has been suggested for the scale factor cutoff measure in Ref.~\cite{DeSimone:2008if}.}  In other words, we replace $\Sigma$ by a new surface $\Sigma_+$, which is the boundary of the future of $\Sigma$.  If $\Sigma$ is spacelike, then $\Sigma_+$ and $\Sigma$ are the same, but if it includes timelike segments, it will be modified in such a way that $\Sigma_+$ is not timelike anywhere.  $\Sigma_+$ will generally include null segments, but these can be made spacelike by an infinitesimal deformation of the surface.  The proposal is that we use $\Sigma_+$ surfaces constructed in this way as our cutoff surfaces.  The corresponding measure can be called the CAH+ measure.  In the following subsections we shall discuss this measure prescription in more detail.

\subsection{Defining the CAH}

We first need to define what is meant by the apparent horizon (AH) in a general spacetime. 
For any spacelike $2D$ surface ${\mathcal S}$ we can construct two null hypersurfaces emanating orthogonally from ${\mathcal S}$ to the future, one corresponding to outward and the other to inward directed light rays.  We shall call ${\mathcal S}$ an AH if the outward going null geodesics are expanding, while the inward going null geodesics have zero expansion (that is, they are neither diverging nor converging as they leave ${\mathcal S}$).\footnote{Note that this definition is different from that given by Bousso in \cite{Bousso}.  We define AH as a $2D$ surface, while Bousso defines it as a $3D$ hypersurface.  Also, his definition refers to a specific observer, and the AH surface depends on the entire observer's worldline.  In contrast, our definition depends only on the local geometry.}\footnote{An apparent horizon defined in this way is a marginally anti-trapped surface.  The region enclosed by such a surface can be thought of as a maximal region whose boundary admits a future-directed light sheet.  The entropy bound (\ref{Nbound}) applies to apparent horizon regions, but generally not to larger regions.}  

The next step is to define CAH.  We start with a smooth segment of spacelike hypersurface $\Sigma_0$, located in an inflating region of some Hubble rate $H_0$ and having 3-curvature $|R^{(3)}|\ll H_0^2$.  We then construct a future-directed, timelike geodesic congruence orthogonal to $\Sigma_0$, labeling the geodesics by their starting points ${\bf x}$ on $\Sigma_0$.  
The scale factor $a({\bf x},\tau)$ can be defined as the cubic root of the volume expansion factor along the geodesic at ${\bf x}$ in a proper time $\tau$, with $\tau=0$ and $a({\bf x},0)=1$ on $\Sigma_0$.  The expansion rate of the congruence is
\beq
H({\bf x},\tau)={\dot a}({\bf x},\tau)/a({\bf x},\tau),
\eeq
where dots stand for derivatives with respect to $\tau$. 

Let us first assume that the spacetime can be locally approximated as FRW, with our geodesic congruence playing the role of comoving geodesics.  This should be a good approximation in inflating regions away from the domain walls and in thermalized regions, as long as effects of structure formation can be neglected.  (Note that geodesics entering a new bubble quickly become comoving in the FRW frame of that bubble \cite{VW}.)   In this case the AH surfaces lying in 3-spaces orthogonal to the congruence are spheres of radius, $r_{AH} = H^{-1}({\bf x},\tau)$.  The comoving apparent horizon radius is then 
\beq
r_{CAH}({\bf x},\tau)=[a({\bf x},\tau)H({\bf x},\tau)]^{-1}={\dot a}^{-1}({\bf x},\tau),
\label{rCAH}
\eeq
and constant CAH surfaces are given by 
\beq
r_{CAH}({\bf x},\tau)= \xi=const.
\label{rxi}
\eeq
Different choices of the surface $\Sigma_0$ should be equivalent; we expect them to be related to large-scale Weyl transformations of the boundary metric (see Sec.~V.A).

In general, AH surfaces will not be spherical, and this leads to an ambiguity in the definition of CAH.  We could, for example, define a constant CAH hypersurface $\Sigma$ by requiring that all AH surfaces on $\Sigma$ enclose the same volume or have the same maximal extent when projected along the geodesic congruence onto the hypersurface $\Sigma_0$.  
At the level of our present understanding we cannot give preference to any of the alternative definitions, but we do not expect these differences to significantly affect the measure.  We shall therefore choose the simplest option and use the prescription of Eqs.~(\ref{rCAH}),(\ref{rxi}) in what follows.  The quantity $r_{CAH}$ in Eq.~(\ref{rCAH}) can be interpreted as the average CAH radius.

\subsection{Properties of cutoff surfaces}

In the regime of inflation with a slowly varying $H$, we have ${\ddot a}\approx H^2 a$, and it follows from Eq.~(\ref{rCAH}) that
\beq
|{\dot r}_{CAH}|\gg |a^{-1}\partial_i r_{CAH}|.
\eeq
This indicates that the surfaces $r_{CAH}=const$ are spacelike and locally nearly parallel to the surfaces $\tau=const$.  However, the character of these surfaces changes drastically as they approach the boundaries of the inflating region.

The inflating region of spacetime is characterized by the condition
\beq
{\ddot a}({\bf x},\tau)>0.
\label{infcond}
\eeq
In this region,
\beq
{\dot r}_{CAH}=-\frac{\ddot a}{{\dot a}^2}<0,
\eeq
so $r_{CAH}$ monotonically decreases along the geodesics.   In the remaining part of spacetime, the expansion decelerates, ${\ddot a}({\bf x}, \tau)< 0$, and we have ${\dot r}_{CAH}>0$.  

Our initial hypersurface $\Sigma_0$, where the geodesic congruence originates, is in the inflating region, so $r_{CAH}$ initially decreases along the geodesics.  If it drops to the value $\xi$, this indicates that we have reached the constant CAH surface $\Sigma$, Eq.(\ref{rxi}).  However, if the geodesic exits the inflating region before reaching $\Sigma$, $r_{CAH}$ begins to grow and will continue to grow until the geodesic either enters another inflating region or terminates at a singularity.  If, for example, a geodesic enters a terminal AdS bubble before reaching $\Sigma$, this geodesic will never cross $\Sigma$ and will end at the big crunch singularity.  This indicates that the hypersurface $\Sigma$ must terminate at the boundary of the inflating region ${\mathcal B}$ (see Fig.~1).

\begin{figure}[t]
\begin{center}
\includegraphics[width=12cm]{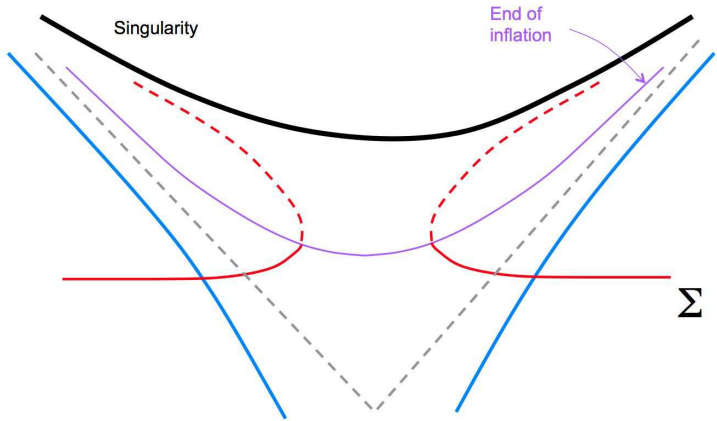}
\caption{Fig.~1. A surface of constant CAH $\Sigma$ for an AdS bubble expanding into a dS parent vacuum. The solid blue line indicates the bubble wall and the dashed grey line is the future light cone of the nucleation event.}
\end{center}
\end{figure}

The geodesics that reached $\Sigma$ before exiting the inflating region can be continued to the future of $\Sigma$.  As we follow such geodesics, $r_{CAH}$ starts from its value $\xi$ on $\Sigma$ and decreases until the geodesic hits the end-of-inflation boundary ${\mathcal B}$.  After that, $r_{CAH}$ starts growing and eventually reaches the value $\xi$ again.  Thus, we can say that $\Sigma$ does not really terminate at ${\mathcal B}$, but gets attached to another sheet $\Sigma'$, shown by a dotted line in Fig.~1.  It is clear that $\Sigma$ becomes timelike near its intersection with ${\mathcal B}$.  Portions of $\Sigma$ in the vicinity of this intersection, and the entire sheet $\Sigma'$ are removed in the modified surface $\Sigma_+$.

Fig.~2 illustrates the cutoff surface $\Sigma_+$ for the same geometry as in Fig.~1.  It includes the spacelike part of the original surface $\Sigma$ and a null portion which is formed by null geodesics emanating from the boundary of the spacelike part of $\Sigma$ (that is, from the 2-surface where $\Sigma$ turns timelike).  Also shown in the figure is the cutoff surface ${\tilde\Sigma}_+$ corresponding to  a smaller value of $\xi$, and thus to a later time.  The null portion of this surface hits the big crunch singularity.  The excluded region is to the future of the cutoff.

\begin{figure}[t]
\begin{center}
\includegraphics[width=12cm]{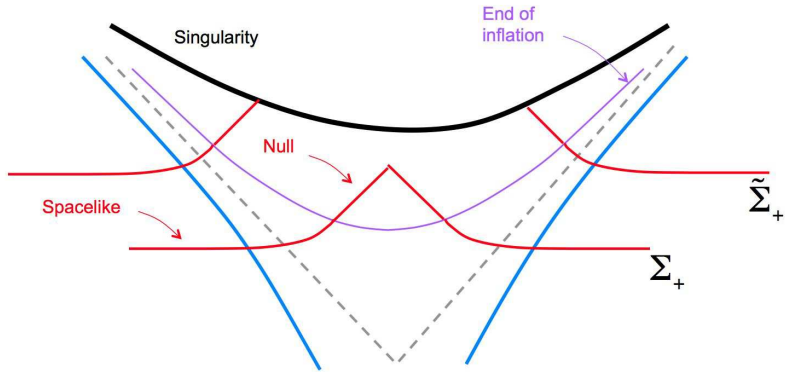}
\caption{Fig.~2.  Cutoff surfaces $\Sigma_+$ for the same spacetime geometry as in Fig.~1.}
\end{center}
\end{figure}

Let us now consider a dS bubble like ours, where inflation is followed by radiation and matter domination, and then by vacuum domination.  The main difference here is that geodesics that exit the inflating region before reaching $\Sigma$  eventually enter the dS vacuum dominated region, where $r_{CAH}$ starts to decrease again, until the value $r_{CAH}=\xi$ is finally reached.
The cutoff surfaces $\Sigma_+$ that intersect the dS bubble at relatively early times will be similar to those for AdS bubbles (see Fig.~3).  However, the later cutoff surfaces will be different and will consist of three distinct components: (i) the spacelike portion of the original surface $\Sigma$ in the parent vacuum, (ii) part of the spacelike portion of the original $\Sigma$ in the vacuum-dominated region inside the bubble, and (iii) a null hypersurface connecting (i) and (ii).

\begin{figure}[t]
\begin{center}
\includegraphics[width=12cm]{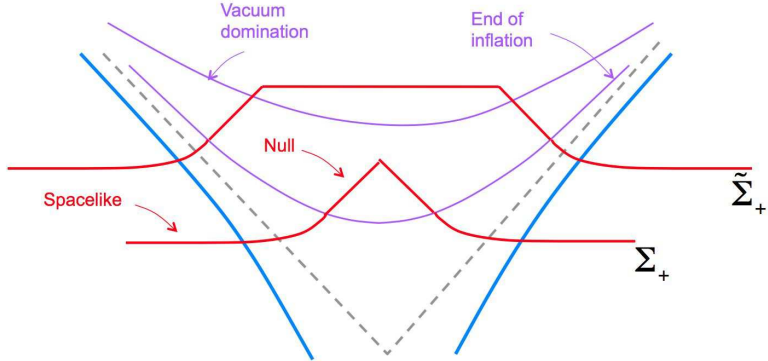}
\caption{Fig.~3.  Cutoff surfaces $\Sigma_+$ for a dS bubble, in which vacuum domination is preceded by radiation and matter dominated epochs. } 
\end{center}
\end{figure}

\section{Discussion}

The picture we have arrived at can be summarized as follows.

The wave function of the multiverse $\Psi_\Sigma[{\bar\phi}]$ is related to the effective action of the boundary theory $W_\xi [{\bar\phi}]$ by the duality
\beq
\Psi_\Sigma = e^{iW_\xi}.
\eeq
In the semiclassical approximation, the 3-spaces $\Sigma$ can be thought of as hypersurfaces in a $4D$ spacetime.  As the boundary cutoff $\xi$ is decreased, these hypersurfaces foliate the spacetime, approaching future infinity.  We suggest that the same hypersurfaces $\Sigma$ should be used as cutoff surfaces for calculating probabilities in the multiverse.

Different choices of the conformal gauge on the boundary correspond to different spacetime foliations and generally yield different probability measures.   We propose that in order to ensure an unbiased sampling of the multiverse, the boundary metric should be fixed by imposing a condition $F=0$, where $F$ is a function only of the metric ${\bar g}_{ij}$ and not of the matter fields.  The simplest choice, that we adopted here, is the condition 
\beq
{\bar R}=0 ,
\label{barR=0}
\eeq
where ${\bar R}$ is the scalar curvature.  We argued that other choices of $F$, satisfying certain consistency requirements, should yield identical probability measures.  

For inflating spacetime regions with a slowly varying expansion rate $H$ (which we called slow-$H$ regions for brevity), the gauge condition (\ref{barR=0}) admits a flat boundary metric, and we found that the cutoff surfaces $\Sigma$ are then the surfaces of constant comoving apparent horizon (CAH).  In parts of spacetime intervening between the slow-$H$ regions, all we can say about the surfaces $\Sigma$ is that they must be spacelike.  

We have introduced a model measure prescription, called CAH+, which has both of these features.  The cutoff surfaces $\Sigma_+$ in this measure can be obtained from the constant $CAH$ surfaces $\Sigma$ by eliminating the future lightcones of all points $P\in \Sigma$.  In other words, $\Sigma_+$ is the boundary of the future of $\Sigma$.  It coincides with $\Sigma$ in slow-$H$ regions and is null (and thus can be made spacelike by an infinitesimal deformation) in the intervening regions.   

The CAH+ measure is somewhat similar to the comoving horizon (CH) measure, which was first introduced in \cite{GSVW} to compare the bubble abundances and later in \cite{Bousso09} as a global spacetime cutoff.  The horizon here is understood to mean the causal horizon.  CH and CAH+ measures coincide in slow-$H$ spacetimes.  In a more general case, however, there are some important differences.

The CH measure has a rather counter-intuitive feature, which can be called "shadows of the future".  The size of the causal horizon on $\Sigma$, and the shape of the surface $\Sigma$ itself, are influenced by the metric in the future of $\Sigma$.  As a result, the probabilities assigned by the CH measure to various events have a non-trivial dependence on the subsequent evolution of the universe.  On the other hand, the apparent horizon depends only on the local geometry, and thus the CAH+ measure does not suffer from shadows of the future.  

The two measures also differ in their predictions.  The phenomenology of CAH+ measure, and in particular its predictions for the cosmological constant, will be discussed elsewhere.

%%%%%%%%%%% %%%%%%%%%%%%%%%%%%%%%%%%%%%%%%%
%%%%%%%%%%%%%%%%%%%%%%%%%%%%%%%%%%%%%%%%%%

\section{Acknowledgements}

I am grateful to Raphael Bousso, Ted Jacobson, Igor Klebanov, Juan Maldacena, Donald Marolf, Delia Schwartz-Perlov, and especially to Jaume Garriga and Ben Freivogel for useful and stimulating discussions.  This work was supported in part by the National Science Foundation (grant PHY-0855447).

\end{document}